\documentclass[conference]{IEEEtran}
\IEEEoverridecommandlockouts

\usepackage{cite}
\usepackage{amsmath,amssymb,amsfonts}
\usepackage{algorithmic}
\usepackage{graphicx}
\usepackage{textcomp}
\usepackage{xcolor}
\usepackage{url}
\usepackage{framed}
\usepackage[hidelinks]{hyperref}  

\def\BibTeX{{\rm B\kern-.05em{\sc i\kern-.025em b}\kern-.08em
    T\kern-.1667em\lower.7ex\hbox{E}\kern-.125emX}}

\begin{document}

\title{How To Mitigate And Defend Against DDoS Attacks In IoT Devices\\}

\author{
\IEEEauthorblockN{Ifiyemi Leigha}
\IEEEauthorblockA{Department of Digital Security and Networks\\
Institut Supérieur d’Électronique de Paris (ISEP)\\
Paris, France\\
ifiyemi.leigha@eleve.isep.fr}
\and
\IEEEauthorblockN{Basak Comlekcioglu}
\IEEEauthorblockA{Department of Electrical and Electronics Engineering\\
Institut Supérieur d’Électronique de Paris (ISEP)\\
Paris, France\\
basak.comlekcioglu@eleve.isep.fr}
\and
\IEEEauthorblockN{María Pilar Bezanilla}
\IEEEauthorblockA{Department of Industrial and Systems Engineering\\
Institut Supérieur d’Électronique de Paris (ISEP)\\
Paris, France\\
maria-pilar.bezanilla-casanueva@eleve.isep.fr}
}

\maketitle


\begin{abstract}
The rapid increase and widespread adoption of the Internet of Things (IoT) across several domains has led to the emergence of new security threats, including Distributed Denial of Service (DDoS). These attacks pose a major concern worldwide because of the significant disruptions they can cause to critical infrastructure and services. IoT devices are vulnerable and attractive to attackers due to their limited security features, making them easy prey for attackers. In addition, attackers can compromise IoT devices to form botnets - a network of private computers infected with malicious software and controlled as a group without the owners' knowledge, e.g., to send spam.
The aim of this paper is to explore different literatures to identify how such DDoS attacks operate in IoT environments and the solutions they provide. Then we will give our idea by proposing a simple theoretical solution that will seek to address this problem.
First, we identify and explain how this attack works, focusing on the Mirai attack as a case study. Next, we will examine the solutions proposed by other authors.
Then, we propose our simple solution reflecting on the challenge from a theoretical perspective. This solution utilizes unique local address mechanisms of the Internet Protocol Version 6 (IPv6) to defend against these attacks.
\end{abstract}

\begin{IEEEkeywords}
Internet of Things, Distributed Denial of Service (DDoS), botnets, IPV6 unique local address, security threat, critical infrastructure and services, security features, malicious software, Mirai botnet.
\end{IEEEkeywords}

\section{Introduction}
In fact, the rapid advancement of the computer network
technology has given rise to the possibility of interconnecting
physical devices such as vehicles, computers, phones, appliances, and other everyday objects embedded with sensors, software, and network connectivity such that these devices can
collect and exchange data with each other and with their
environment [1]. This network of interconnected devices is 
called the Internet of Things (IoT). The problem is that these IoT
devices have some security lapses such as inadequate security
measures (improper authentication mechanisms, encryption), 
default or weak credentials, lack of firmware updates, large
attack surface, bandwidth amplification, limited processing
power and memory and lack of user awareness [2]. 
These weaknesses have paved the way for attackers to
attack these devices. One of the most notable attacks on these
devices used by attackers is the distributed denial of service
(DDoS) attack. DDoS is a coordinated attack in which an
attacker sends commands through a command and control
(C\&C) server to zombie agents (botnets - infected IoT
devices) so that they can perform malicious activities such as
flooding a server [4]. An example of such attack is the Mirai 
botnet attack which took down the infrastructure of Dyn, a 
major Domain Name System (DNS) provider. As a result, 
popular websites like Netflix, Twitter and so on experienced 
massive disruptions. It implemented the attack with an estimated amount between 100,000 – 200,000 botnets [5]. The 
three types of DDoS attacks include volumetric attacks, 
protocol attacks and application layer attacks. So many 
defense mechanisms and strategies have been identified by 
researchers which can be used to defend and mitigate against 
this type of attack. Such strategies include Software-Defined 
Networks (SDN) [9] and Edge Computing [6]. In the 
subsequent sections, we highlight these strategies and as 
well reflect on this challenge and propose a theoretical 
solution. Our solution involves employing IPv6 Unique Local 
Addressing (ULA) as a means of communication among the 
connected devices. This strategy proves effective because devices can communicate using this method without having to go through the Global Internet, which is really the 
propeller of DDoS attacks. Our solution can contribute greatly 
to the defense mechanism by adding an additional layer of 
network isolation and reducing the attack surface. Our 
solution proposes the use of multiple defense strategies, with 
ULA, as the core communication means in the network. We 
segment the network of IoT devices with ULAs based on 
device type and we implement several other strategies like 
Firewalls and Access Control List (ACL), ingress and egress 
filtering at the network edge devices (routers and firewalls), 
rate limiting and traffic shaping strategies, intrusion detection 
and prevention systems (IDS/IPS) and continuous monitoring and incidence reporting.

\section{The Mirai Botnet Attack}
Mirai is a malware which infected several vulnerable IoT devices and turned them into bots that could be used for Distributed Denial of Service (DDoS) attacks[10]. This attack called the Mirai botnet attack occurred in 2016 and it affected millions of devices connected to the Internet of Things[10]. One thing about these targeted devices such as routers, cameras, and other connected devices is that they are either protected by default passwords or are running unpatched software among others [10]. The malware was designed to scan the Internet for vulnerable devices and then infect them with the Mirai bot. Once a device was infected, it became part of the network of bots that could be used to launch DDoS attacks. This attack was one of the largest DDoS attacks ever recorded, with a peak traffic rate of 1.2 terabits per second. The attack targeted Dyn, a major DNS provider, and disrupted access to major websites such as Amazon, Twitter, etc. [10]. This attack disrupted online services, resulted in economic loss, raised public safety concerns, and caused reputation damages [10].

\begin{figure}[htbp]
    \centering
    \includegraphics[width=0.45\textwidth]{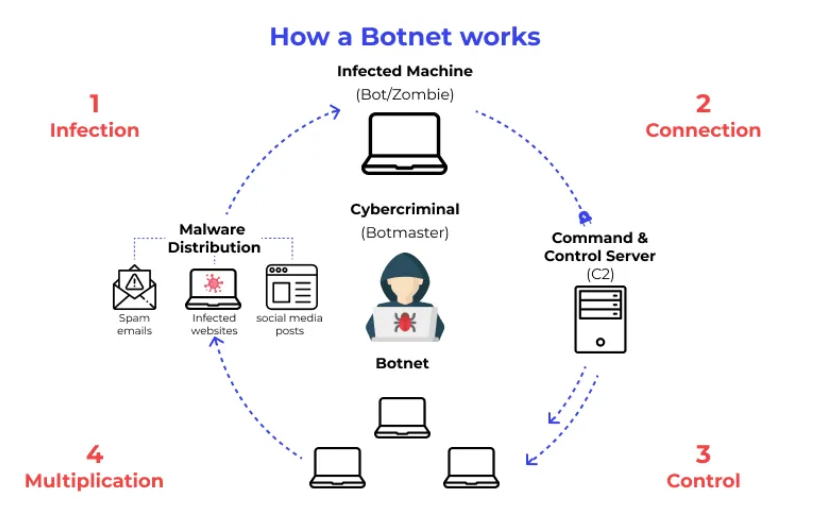}
    \caption{How a Botnet Works – DDoS Lifecycle. Image adapted from~\cite{b10}.}
    \label{fig:ddosbotnet}
\end{figure}

\subsection{How Botnets Work}
Figure 1 shows how a botnet work. In stage 1, cyber criminals infect machines/distribute malware via spam emails, infected websites or social media post. Infected devices become bots in the botnet. In step 2, the infected devices contact the attacker's command and control center (C2). The botmaster sends commands to the bots via the C2 server in step 3. The infected bots start spreading the malware further in step 4.

\section{LITERATURE REVIEW}
Distributed denial-of-service (DDoS) attacks pose a 
significant threat to the availability and integrity of online 
services and networks. These attacks overwhelm the target 
system with a massive volume of malicious traffic, rendering 
it unable to handle legitimate user requests. This literature 
review aims to explore the solutions on how to mitigate and 
defend against DDoS attacks already presented by other 
researchers in their papers. According to [8], machine learning based detection framework can be used to predict the possibility of an 
abnormal activity based on a log file generated by a honey pot, 
using a light weighted classification algorithm, preferably an 
unsupervised one. In this case, a honey pot is intentionally 
used to lure in attackers who will attempt to inject malware 
into the system through an open port say Telnet port 23 or 
2323. The purpose of this is to capture the malware properties 
and its style of invading the security of IoT devices. This log 
file can contain information such as new malware families and 
their variants, type of targeted devices, server IP address, port 
numbers etc. These information on the log files are 
transformed into a proper table format that will work as data 
sets so that it can be used to train their machine learning 
model, which in turn is implemented in the network to detect 
traffic patterns like the data they were trained with. [7] takes 
the approach of leveraging computational resources at the 
edge of the network to accelerate the defense from IoT-DDoS 
attacks and arrest them before they can cause considerable 
damage. They propose ShadowNet - an architecture that 
makes the edge the first line of defense against IoT-DDoS. [3] 
proposed building the IoT architecture as a Software-Defined 
Network (SDN)-based traffic monitoring and anomaly 
detection framework so that since IoT devices normally have 
reasonably predictable traffic pattern during normal 
operations, if there is any anomaly, it can be detected. 
Typical components of such systems include SDN (Software 
Defined Networking) controllers, switch and IoT devices. 
This system setup tries to learn the IoT device’s normal patterns to block communication that is out of the ordinary.

\section{OUR MODEL}
Our model involves two key processes: 
A. Segmentation of IoT devices using IPv6 unique 
local addresses (ULAs) 
 IPv6 unique local addresses (ULAs) help devices in a 
local private network to communicate securely. ULAs are 
not reachable from the Global Internet. So, in this model, we 
divide our network into different segments based on certain 
criteria like device type and assign ULAs to each segment. 
No matter how large the private network is, ULAs can help 
departments, sites and so on to communicate securely. 
B. Implement the edge paradigm at the network 
perimeter 
 At the network perimeter, we implement access control 
and filtering mechanisms. This can include firewalls, 
intrusion detection systems, intrusion prevention systems or 
access control lists that monitor and filter traffic entering or 
leaving the private IoT network. We set up ingress and 
egress filtering mechanisms, implement rate limiting and 
traffic shaping mechanisms to control the flow of traffic to 
and from the IoT segments. We also set up appropriate 
thresholds to limit the maximum number of connections, 
packets per second, or bandwidth allocated to each segment. 
In Fig 1 below., the Gateway which is also a firewall acts as 
the network perimeter and provides access control and 
filtering capabilities. The IoT devices are assigned unique 
IPv6 ULAs and communicate with each other within the 
private network. The firewall monitors the traffic and 
applies security measures to mitigate DDoS attacks. 
Although this work proposes a theoretical model, the described mechanisms can be implemented in practice using modern network simulation tools. For instance, the segmentation of IoT devices using IPv6 ULAs can be configured on virtual routers using platforms like GNS3 or Cisco Packet Tracer. Rate limiting and ingress/egress filtering policies can be enforced using Access Control Lists (ACLs) and Quality of Service (QoS) features. The firewall policies at the gateway can be simulated using open-source tools such as pfSense or iptables. A potential testing setup could involve traffic generators simulating DDoS behavior, while the impact of our model would be evaluated by monitoring device isolation, traffic shaping, and drop rates. Future work can involve setting up such simulations to quantify the efficiency of the proposed model.

\begin{figure}[htbp]
    \centering
    \includegraphics[width=0.48\textwidth]{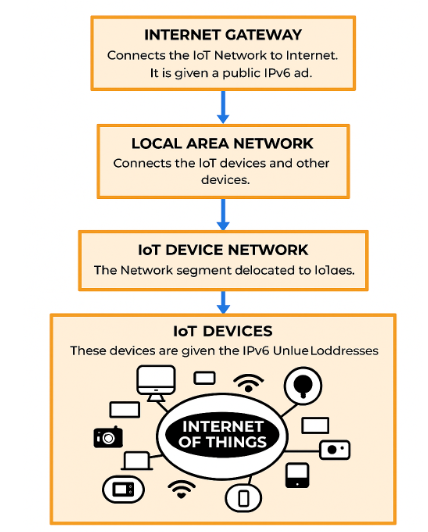}
    \caption{IPv6 Unique Local Address (ULA) allocation in IoT networks. }
    \label{fig:ipv6-ula-iot}
\end{figure}
This model segments IoT devices into private communication groups and connects them via an Internet Gateway, with ULA-enforced isolation.

\section{CONCLUSION}
In conclusion, by using IPv6 ULAs and implementing 
appropriate security measures, the IoT environment can 
benefit from enhanced security and isolation, reducing the 
impact of potential DDoS attacks. Researchers, industries, and academicians can conduct further research on these concepts.

\vspace{12pt}


\begin{thebibliography}{00}

\bibitem{b1} M. Alshamkhany, W. Alshamkhany, M. Mansour, M. Khan, S. Dhou, and F. Aloul, “Botnet attack detection using Machine Learning,” in *Proc. 14th Int. Conf. Innovations in Information Technology (IIT)*, Al Ain, United Arab Emirates, 2020, pp. 203–208.

\bibitem{b2} M. H. Rohit, S. M. Fahim, and A. H. A. Khan, “Mitigating and detecting DDoS attack on IoT Environment,” in *Proc. IEEE Int. Conf. Robotics, Automation, Artificial-Intelligence and Internet-of-Things (RAAICON)*, Dhaka, Bangladesh, 2019, pp. 5–8.

\bibitem{b3} K. M. S. Azad, N. Hossain, M. J. Islam, A. Rahman, and S. Kabir, “Preventive determination and avoidance of DDoS Attack with SDN over the IoT Networks,” in *Proc. Int. Conf. Automation, Control and Mechatronics for Industry 4.0 (ACMI)*, Rajshahi, Bangladesh, 2021, pp. 1–6.

\bibitem{b4} K. Al-Begain, et al., “A DDoS detection and prevention system for IoT devices and its application to Smart Home environment,” *Applied Sciences*, vol. 12, no. 22, p. 11853, Nov. 2022.

\bibitem{b5} Z. Ahmed, S. M. Danish, H. K. Qureshi, and M. Lestas, “Protecting IoTs from Mirai Botnet attacks using Blockchains,” in *Proc. IEEE 24th Int. Workshop on Computer Aided Modeling and Design of Communication Links and Networks (CAMAD)*, Limassol, Cyprus, 2019, pp. 1–6.

\bibitem{b6} S. Bhatia, et al., “Distributed denial of service attacks and defense mechanisms: Current landscape and future directions,” in *Advances in Information Security*, Springer, 2018, pp. 55–97.

\bibitem{b7} K. Bhardwaj, J. C. Miranda, and A. Gavrilovska, “Towards IoT-DDoS prevention using edge computing,” in *Proc. Workshop on Hot Topics in Edge Computing (HotEdge)*, 2018.

\bibitem{b8} R. Vishwakarma and A. K. Jain, “A honeypot with machine learning based detection framework for defending IoT-based botnet DDoS attacks,” in *Proc. 3rd Int. Conf. Trends in Electronics and Informatics (ICOEI)*, Tirunelveli, India, 2019, pp. 1019–1024.

\bibitem{b9} H. Jing and J. Wang, “Detection of DDoS attack within industrial IoT devices based on clustering and graph structure features,” *Security and Communication Networks*, vol. 2022, Hindawi, Mar. 2022, pp. 1–9.

\bibitem{b10} D. C. Subudhi, “A Case Study on Mirai Botnet Attack of 2016,” *Medium*, May 2023. [Online]. Available: \url{https://medium.com/@d21dcs151/a-case-study-on-mirai-botnet-attack-of-2016-4b66630e6508}

\end{thebibliography}
\end{document}